\documentclass{article}
\usepackage{spconf,amsmath,graphicx}
\usepackage{booktabs}
\usepackage{amsfonts}
\usepackage{multirow}
\usepackage{booktabs}
\usepackage{bbding}
\usepackage{pifont}
\usepackage{algorithm}
\usepackage{algpseudocode}
\usepackage{color}
\usepackage{xcolor}
\usepackage{caption}
\usepackage{url}

\title{Improving Speaker Diarization Using Semantic Information: Joint Pairwise Constraints Propagation}
\name{Luyao Cheng, Siqi Zheng, Qinglin Zhang, Hui Wang, Yafeng Chen, Qian Chen, Shiliang Zhang}
\address{Speech Lab, Alibaba Group \\
\tt \normalsize\{shuli.cly, zsq174630, tanqing.cq, sly.zsl\}@alibaba-inc.com}
\begin{document}
%
\maketitle

\begin{abstract}

Speaker diarization has gained considerable attention within speech processing research community. Mainstream speaker diarization rely primarily on speakers' voice characteristics extracted from acoustic signals and often overlook the potential of semantic information. Considering the fact that speech signals can efficiently convey the content of a speech, it is of our interest to fully exploit these semantic cues utilizing language models. In this work we propose a novel approach to effectively leverage semantic information in clustering-based speaker diarization systems. Firstly, we introduce spoken language understanding modules to extract speaker-related semantic information and utilize this information to construct pairwise constraints. Secondly, we present a novel framework to integrate these constraints into the speaker diarization pipeline, enhancing the performance of the entire system. Extensive experiments conducted on a public dataset demonstrate the consistent superiority of our proposed approach over acoustic-only speaker diarization systems.

\end{abstract}


\section{Introduction}
\label{sec:introduction}
Speaker Diarization(SD) is the task of solving the question ``who speak when" and assigning the speaker labels for the given audio. In most applications setting, the speaker label will be integrated with the corresponding words or sentences transcribed from Automatic Speech Recognition(ASR) system. Despite the rich profusion of transcribed text, mainstream SD systems\cite{Park2021ARO} take only acoustic information into consideration. A traditional SD system usually consists of the following components: (1) A voice activity detection(VAD) component. (2) A speaker embedding extractor, such as x-vector\cite{Snyder2018XVectorsRD}, d-vector\cite{Wan2017GeneralizedEL} and ECAPA-TDNN\cite{Dawalatabad2021ECAPATDNNEF}. (3) A speaker clustering component using clustering algorithms such as agglomerative hierarchical clustering(AHC)\cite{Han2008StrategiesTI} and spectral clustering(SC)\cite{Wang2017SpeakerDW}. 
Mainstream SD system solely utilize acoustic information, ignoring the potential of content semantics. This limitation often results in obvious performance degradation in adverse acoustic conditions such as noise, reverberation, and far-field recordings. 

Some previous works tried to leverage semantic information by learning speaker-role information in specific domain applications such as \textit{air traffic controller}\cite{ZuluagaGmez2021BertrafficBJ} or \textit{medical consultation}\cite{Flemotomos2022MultimodalCW}. However, these methods are highly task-oriented and only suitable for two-speaker scenarios. In this work, we focus on open multi-party meeting scenarios where the number of speakers is unknown 
and the relations among speakers are unspecified. 

Recent works such as \cite{Kanda2021TranscribetoDiarizeNS} and \cite{Xia2021TurntoDiarizeOS} tried to implicitly utilize semantic information with an modified ASR system. These systems often require large-scale annotated multi-speaker speech data, which is scarce in reality and extremely expensive to obtain. Furthermore, these works mostly utilize semantic information to precisely determine the turning point between speaker utterances. Semantic information is not explicitly used in speaker clustering and determining the number of speakers. Other multi-modal speaker diarization systems presented similar limitations \cite{Khare2022ASRAwareEN}\cite{Park2018MultimodalSS}\cite{Paturi2023LexicalSE}.

To address these limitations, we explicitly incorporates semantic information into speaker embedding normalization and speaker clustering. We introduced additional spoken language processing (SLP) modules to extract speaker-related information from transcribed texts. The main contributions of this paper are as follows:
(1) We propose a novel framework to directly incorporate semantic information into a speaker clustering, exceedingly the performance boundary of acoustic-only speaker clustering.

(2) We introduce methods of pairwise constraints propagation to speaker clustering, and investigate the effectiveness of constraints derived from semantic information.

\vspace{-5px}

\section{Semantic speaker constraints}
\label{sec:proposed_methods}

\subsection{Semantic Speaker-related Tasks}
For the given conversation speech signal features $S = \{s_1, s_2, ..., s_T\}$, the text contents $Y = \{y_1, y_2, ..., y_K\}$ are decoded by the ASR system from $S$. In traditional speaker diarization system, we extract embeddings $E = \{e_1, e_2, ..., e_{N}\}$ from the speech signal $S$ by speaker embedding extractor. In most application settings, A forced alignment(FA) module is utilized to associate speech signal features $S$, transcribed text $T$ and speaker embeddings $E$. This cascaded pipeline often results in cumulative errors, as each component is optimized for different objectives. 

We defined two SLP tasks: \textbf{Dialogue-Detection} and \textbf{Speaker-Turn-Detection} to extract speaker-related information based on the transcribed text $T$. Due to the uncertainty of the actual duration of the meetings, in practical applications, the task should be defined on the subsequence from the whole session.

\textbf{Dialogue-Detection} takes a sequence of sentences as input and predicts whether this is transcribed from multi-speaker dialogue or single-speaker. Dialogue-detection can be defined as a binary text classification problem.

\textbf{Speaker-Turn-Detection} tries to determine, for each given sentence in the sequence, the probability of the occurrence of speaker change. Speaker turn detection can be defined as a sequence labeling problem, where the goal is to determine whether the given position represents a point of change in speaker role from a semantic perspective. 

In practice, ASR system introduced insert, delete and replacement errors into the transcribed text which will cause the performance of SLP tasks decreasing. In \cite{Cheng2023ExploringSI}, a simple yet effective hybrid strategy was proposed to mitigate the impact of ASR system errors by incorporating both acoustic and semantic information. In this paper, we extended this approach to improve the accuracy of these two SLP models.

\subsection{Pairwise Constraints from Semantic Information}
\label{ssec:build_pc}

Traditional cluster-based SD systems often employ Text-Independent speaker embedding models. Consequently, extracting speaker information from semantics during the embedding extraction stage becomes challenging. We proposed that utilizing semantic information to construct constraints between embeddings offers a suitable strategy for incorporating the results of SLP models into cluster-based SD.

We construct two kinds of constraints from the semantic speaker-related information: must-link $\mathcal{M}$ and cannot-link $\mathcal{C}$:
\begin{equation}
\begin{aligned}
    \mathcal{M} &= \{(e_i, e_j): l(e_i) = l(e_j)\} \\
    \mathcal{C} &= \{(e_i, e_j): l(e_i) \neq l(e_j)\}
\end{aligned}
\end{equation}
where the $l(\cdotp)$ means the speaker role of the embedding. 

\begin{figure}[ht!]
    \centering
    \includegraphics[width=0.475\textwidth]{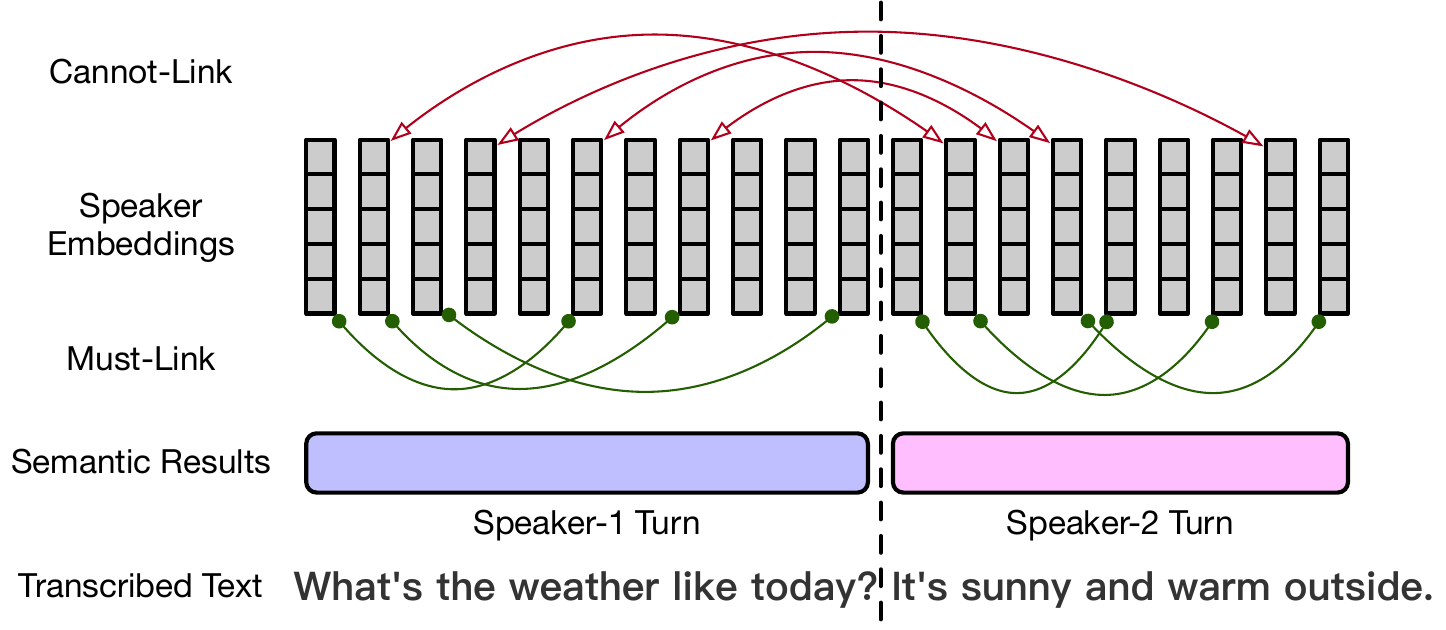}
     \caption{A sample of strategy for constructing constraints. }
      \vspace{-12px}
    \label{fig:pcc_constraints}
\end{figure}
As shown in Figure \ref{fig:pcc_constraints}, the strategy of building $\mathcal{M}$ and $\mathcal{C}$ can be concluded as: If two embeddings contained in one non-dialogue segments, a must-link between two embeddings should be constructed. If two embeddings cross speaker-turn change point, a cannot-link should be constructed.

\begin{figure*}[ht!]
    \centering
    \includegraphics[width=0.8\textwidth]{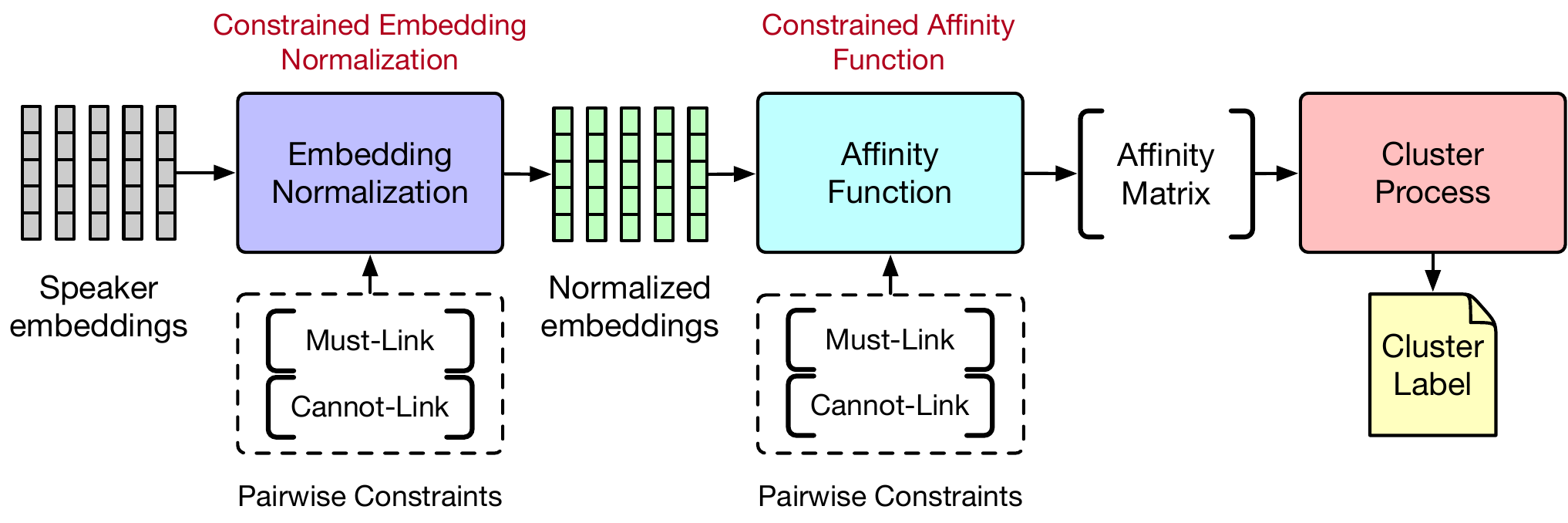}
    \captionsetup{skip=2.5pt}
    \caption{The pipeline is a traditional speaker diarization backend with acoustic information. The addtional pairwise constraints constructed from semantic information, including Must-Link and Cannot-Link, will be used in two parts: Embedding Normalization and Affinity Function. }
    \vspace{-16px}
    \label{fig:pcc}
\end{figure*}

\vspace{-5px}

\section{Constrained Speaker Diarization}

\vspace{-5px}

We proposed a novel framework named \textbf{Joint Pairwise Constraints Propagation (JPCP)}, as shown in Figure \ref{fig:pcc}. The proposed framework incorporates pairwise constraints into embedding normalization and affinity function. The following sections provide more details of these components.

\subsection{Constrained Embedding Normalization}
\label{ssec:constrained_embedding_normalization}

We introduced the semi-supervised dimension reduction (SSDR) algorithm\cite{Zhang2007SemiSupervisedDR} to integrate pairwise constraints into speaker embedding normalization module.

We first build a weight matrix $\mathbf{S}$ from pairwise constraints:
\begin{equation}
    \label{eq:SSDR_weight}
    \mathbf{S}_{ij} = \begin{cases}
        \frac{1}{N^2} + \frac{\alpha}{|\mathcal{M}|} & \text{ if } (e_i, e_j) \in \mathcal{M} \\
        \frac{1}{N^2} - \frac{\beta}{|\mathcal{C}|} & \text{ if } (e_i, e_j) \in \mathcal{C} \\
        \frac{1}{N^2} & \text{ otherwise }
    \end{cases}
\end{equation}
The function of constrained embedding normalization is to find the projective vectors $\mathbf{W} = [\mathbf{w}_1, \mathbf{w}_2, ..., \mathbf{w}_d]$, such that the normalized (low-dimensional) embeddings $\{\mathbf{w}^{T}e_k\}$ can preserve the manifold of the original embeddings as well as the pairwise constraints sets $\mathcal{M}$ and $\mathcal{C}$. The objective function can be defined as $J(\mathbf{W})$:

\begin{equation}
    \label{eq:SSDR_objective}
    J(\mathbf{W}) = \mathbf{W}^{T}E\mathbf{L}E^{T}\mathbf{W}    
\end{equation}
where $\alpha$, $\beta$ are parameters to balance the contribution of each constraint type. The $\mathbf{L}$ is the Laplacian matrix of the weight matrix $\mathbf{S}$. The problem expressed by \eqref{eq:SSDR_objective} is a typical eigen-problem and can be efficiently solved by computing the eigenvectors of $E\mathbf{L}E^{T}$ corresponding to the largest eigenvalues.

\subsection{Constrained Affinity Function}
\label{ssec:constrained_affinity_propagation}
Constructing the affinity matrix $\mathcal{A} = \{\mathcal{A}_{ij}\}_{N \times N}$ is another core part of speaker clustering algorithms, especially in spectral clustering. In \cite{Wang2017SpeakerDW}, a series of refinement operations are defined to refine the affinity matrix, like row-wise thresholding and symmetrization.

The constrained affinity function can be seen as a new refinement operation which integrates with $\mathcal{M}$ and $\mathcal{C}$. The constraints are concluded into a constraints matrix $\mathcal{Z}$:
\begin{equation}
    \mathcal{Z}_{ij} = \begin{cases}
        +1 & \text{ if } (e_i, e_j) \in \mathcal{M} \\
        -1 & \text{ if } (e_i, e_j) \in \mathcal{C} \\
        0 & \text{ otherwise }
    \end{cases}
\end{equation}

Since pairwise constraints cannot cover all the pairs $(i, j)$ in the affinity matrix, a \textbf{constraint propagation algorithm} is introduced to build the global constraints relationship. The propagated pairwise constraints matrix $\hat{\mathcal{Z}} = f(\mathcal{Z})$, and the $f(\cdotp)$ is the constraint propagation function. 
The propagated pairwise constraints matrix $\hat{\mathcal{Z}}$ and result affinity matrix $\hat{\mathcal{A}} \in \mathcal{R}^{N \times N}$ should satisfy:

\begin{equation}
    \hat{\mathcal{A}}_{ij} = \begin{cases}
        1 - (1 - \hat{\mathcal{Z}}_{ij})(1 - A_{ij}) & \text{ if } \hat{\mathcal{Z}}_{ij} \ge 0 \\
        (1 + \hat{\mathcal{Z}}_{ij}) A_{ij} & \text{ if } \hat{\mathcal{Z}}_{ij} < 0
    \end{cases}
\end{equation}

In practice, a classical constraints propagation methods called $\text{E}^{2}\text{CP}$ \cite{Lu2011ExhaustiveAE} has been applied:
\begin{equation}
\label{eq:e2cp}
    \hat{\mathcal{Z}} = (1 - \lambda)^{2}(\mathbf{I} - \lambda\mathbf{L})^{-1}\mathcal{Z}(\mathbf{I} - \lambda\mathbf{L})^{-1}
\end{equation}
where $\mathbf{L} = \mathbf{D}^{-1/2}\mathcal{A}\mathbf{D}^{-1/2}$ is the Normalized Laplacian matrix and $\mathbf{D}$ is the degree matrix of $\mathcal{A}$. The parameter $\lambda \in [0, 1]$ is to control the effectiveness of the constraints matrix.

\subsection{Improve Constraints Propagation Algorithm}
\label{ssec:E2CPM}
The constraints derived from semantic information have certain limitations: (1) incorrect constraints may be generated when the semantic model makes prediction errors, and (2) the constraints may become too close to the embedding when there are frequent speaker-turn changes. To address these issues, we propose an enhanced version of the $\text{E}^{2}\text{CP}$ method, referred to as $\text{E}^{2}\text{CPM}$:

Firstly, we introduce a k-NN strategy for constructing the Laplacian matrix $\mathbf{L}$ in equation \eqref{eq:e2cp}. Specifically, we defined the affinity matrix $\mathcal{A}' = \{a'_{ij}\}_{N \times N}$ as follows: $a'_{ij} = a_{ij}$ if $e_j (j \neq i)$ is among the k-nearest neighbors of $e_i$ and $a'_{ij} = 0$ otherwise. To ensure symmetry, we set $\mathcal{A}'' = (\mathcal{A}'^{T} + \mathcal{A}') / 2$.

Secondly, we enhance the existing constraints by incorporating embedding pairs with a high confidence level in affinity similarity. Specifically, we randomly select pairs with affinity scores above a threshold $\theta_m$ and add them as additional must-link constraints $\mathcal{M}' = \mathcal{M} \cup r_{\theta_m}(\mathcal{A})$. Similarly, we randomly select pairs with affinity scores below a threshold $\theta_c$ as additional cannot-link constraints $\mathcal{C}' = \mathcal{C} \cup r_{\theta_c}(\mathcal{A})$ where $r_{\theta_m}(\cdot)$ and $r_{\theta_c}(\cdot)$ represent thresholding and uniform random functions, respectively.

\vspace{-5px}

\section{Experimental setup}
\label{sec:experimental_setup}

\subsection{Dataset and Metrics}
Our experiments are conducted on AISHELL-4 \cite{Fu2021AISHELL4AO} 
which focuses on multi-party meeting scenario, where all speech content is manually annotated. 

\begin{table*}[ht!]
\centering
\resizebox{\textwidth}{!}{%
\begin{tabular}{cccccccc}
\hline
\multirow{2}{*}{Diarization System}                                                         & \multirow{2}{*}{Constraints} & \multirow{2}{*}{Methods} & \multicolumn{3}{c}{Cluster Metrics} & \multicolumn{2}{c}{Diarization Metrics} \\ \cline{4-8} 
                                                                                            &                              &                                      & ARI       & NMI      & SpkDiff \#   & CpWER (\%)        & TextDER (\%)        \\ \hline
Acoustic Only                                                                        & No Constraints               & SC                                   &   -  &  -  &      -      &     26.1816       &       3.7723        \\ \hline
Semantic Turn-Cut                                                                    & No Constraints               & SC                                   &   0.8901  &  0.8616  &      11      &     25.6421      &       3.4636        \\ \hline
\multirow{4}{*}{\begin{tabular}[c]{@{}c@{}}JPCP-I\end{tabular}}   & Inference Constraints         & SSDR + SC                            &    0.9010       &      0.8863    &     11         &        25.9185           &    3.8122                 \\
                                                                                            & Inference Constraints         & $\text{E}^{2}\text{CP}$                 &     0.9006    &    0.8857      &     11         &       25.9174            &      3.8161               \\
                                                                                            & Inference Constraints         & $\text{E}^{2}\text{CPM}$                &      0.9162      &    0.8863      &     10         &     25.2774             &    3.0967                 \\
                                                                                            & Inference Constraints         & SSDR + $\text{E}^{2}\text{CPM}$         &      \textbf{0.9171}  & \textbf{0.8871}  &   \textbf{9}  &  \textbf{25.3168}   &     \textbf{3.0379}   \\ \hline
\multirow{2}{*}{\begin{tabular}[c]{@{}c@{}}JPCP-S \end{tabular}} & Simulation 6\%    & SSDR + $\text{E}^{2}\text{CPM}$                         &   0.9939  &  0.9879  &      4       &     24.5919       &         1.9810       \\
                                                                                            & Simulation 12\%   & SSDR + $\text{E}^{2}\text{CPM}$        &   \textbf{0.9961}  &  \textbf{0.9927}  &   \textbf{3}  &    \textbf{24.4809}      &       \textbf{1.9028}       \\ \hline
\end{tabular}%
}
\caption{Performance evaluation of cluster metrics and speaker diarization results. TextDER refers to the amount of text assigned to wrong speakers. SpkDiff \# refers to difference in number of speakers between inference and ground truth. JPCP-S utilizes ground truth ASR results, indicating potential upper bound for our proposed system.}
\vspace{-12px}
\label{tab:pcc_results}
\end{table*}

We report the following clustering algorithm metrics: Normalized Mutual Information (NMI) and Adjusted Rand Index (ARI).
As the transcribed text and forced-alignment module have been used in the pipeline, we directly report the Concatenated Minimum-permutation Word Error Rate (cpWER). Additionally, we use the metric Text Diarization Error Rate (TextDER), to evaluate the amount of text assigned to wrong speakers \cite{Cheng2023ExploringSI}.

\subsection{Acoustic and Semantic Modules Configuration}
The whole system pipeline we utilized in this paper is similar with the pipelines in \cite{Cheng2023ExploringSI} and we improved some acoustic models. The speaker embedding extractor is based on CAM++ \cite{Wang2023CAMAF} trained on a large Mandarin corpus\footnote{The speaker embedding extractor we used can be found in \url{https://github.com/alibaba-damo-academy/3D-Speaker} }. The ASR model we used is based on Paraformer\cite{Gao2022ParaformerFA} trained by FunASR \cite{Gao2023FunASRAF} toolkits\footnote{The ASR models and punctuation prediction models we used can be found in \url{https://github.com/alibaba-damo-academy/FunASR} }. These models are open-sourced and fixed in all our experiments. 

The semantic models trained for Dialogue-Detection and Speaker-Turn-Detection tasks are based on the pre-trained BERT language model. The training samples are generated by a sliding-window method with a window length of 64 and a shift of 16 and the training labels for these two semantic tasks can be obtained by the manually annotated speaker label from the speech content.

The construction of pairwise constraints using semantic information is explained in Section \ref{ssec:build_pc}. To explore the efficacy of our proposed method, we also generated simulated pairwise constraints. These simulated constraints were created from the ground truth speaker labels assigned to each embedding. We randomly selected a subset of pairs from all the possible embedding pairs.

\vspace{-5px}

\section{Results and Disscussions}
\label{sec:results}

\vspace{-5px}

\subsection{Experiments Results}
The experiments results are shown in the Table \ref{tab:pcc_results}.
The baseline is the acoustic speaker diarization system that combines VAD, CAM++ and SC. The ``Semantic Turn-Cut'' settings, proposed from \cite{Cheng2023ExploringSI}, optimized the strategy for segmenting embeddings by integrating the timestamps of semantic boundaries into the VAD results. The following JPCP experiment also utilized this approach to extract embeddings.

We report the results of incorporating simulation constraints (JPCP-S) and inference constraints (JPCP-I). JPCP-S
utilizes ground truth ASR results and directly simulates constraints for speaker clustering, indicating the potential upper bound for our proposed system. The simulation constraints demonstrates a significant improvement in speaker diarization, particularly in determining the number of speakers. 

In terms of inference constraints, it can be observed that the improvement achieved by SSDR is relatively smaller compared to $\text{E}^{2}\text{CPM}$. Hence, it is inferred that the constrained affinity function has a more direct impact on the overall clustering. Our proposed JPCP approach has shown improvements compared to the baseline, with a $19\%$ decrease in TextDER, and also some improvement in SpkDiff. It can also be observed that both methods are sensitive to the quality of constraints so that the improvement achieved by JPCP-I in the experimental results is relatively marginal.

\subsection{Constraints Analysis}

\begin{figure}[ht!]
    \centering
    \includegraphics[width=0.475\textwidth]{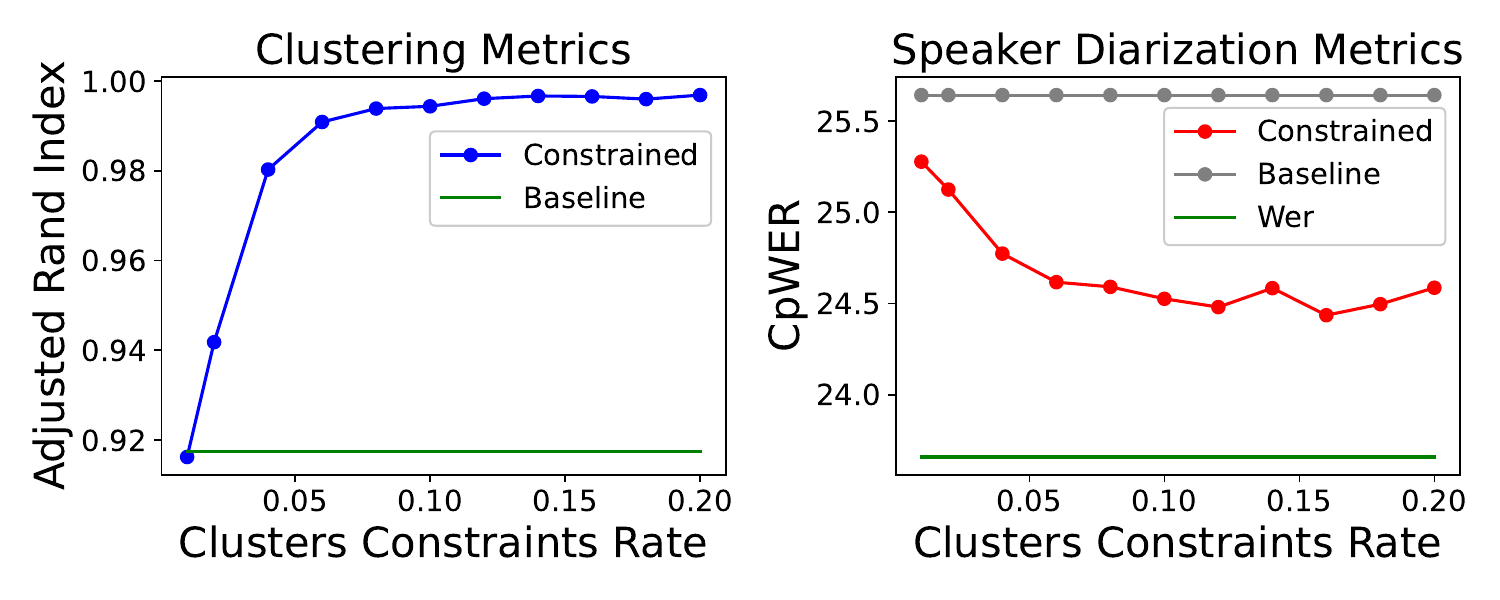}
     \caption{The impact of pairwise constraints rate on both clus- tering metrics and the effectiveness of the overall speaker diarization system.}
     \vspace{-18px}
    \label{fig:pcc_tendency_analysis}
\end{figure}

Given the flourishing development of large-scale language models, the semantic constraints constructed through semantic information are becoming increasingly robust. Therefore, in this paper, we simulate higher-quality constraints to explore the performance upper bound of our proposed methods. 

Figure \ref{fig:pcc_tendency_analysis} shows that with the number of constraints increases, both the clustering performance and the effectiveness of speaker diarization show significant improvements. It can be observed that with around $6\%$ of constraints, the results approach the system's upper bound. This indicates the high potential of our proposed method.

\vspace{-5px}

\section{Conclusion}
\label{sec:conclusion}

We propose a novel architecture that integrates semantic modeling into a clustering-based speaker diarization system, enhancing its overall performance. Speaker-related information is extracted from ASR transcriptions and represented as pairwise constraints. We investigate the integration of these constraints in the process of speaker embedding normalization and the speaker affinity function. Experimental results show that incorporating semantic constraints improves performance compared to acoustic-only models. Moreover, our system architecture is designed to be compatible with other modules and shows promising results in simulated experiments. As the current work has demonstrated the potential of our framework, our future work will focus on further enhancing the quantity and quality of pairwise constraints to achieve superior results.

\vfill\pagebreak


\bibliographystyle{IEEEbib}
\bibliography{main}

\end{document}